\documentstyle[prl,aps,preprint]{revtex}
\begin{document}
\title{The underscreened Kondo effect: a two S=1 impurity model}               
\author{Karyn Le Hur and B. Coqblin}
\address{Laboratoire de Physique des Solides, Universit{\'e} Paris--Sud,
		    B{\^a}t. 510, 91405 Orsay, France}
 \maketitle
 \begin{abstract}
 The underscreened Kondo effect is studied within a model of two impurities S=1
 interacting with the conduction band and via an interimpurity coupling
 $K\vec{S_1}.\vec{S_2}$. Using a mean-field treatment of the bosonized Hamiltonian, we
 show that there is no phase transition, but a continuous cross-over versus K
 from a non Kondo behaviour to an underscreened Kondo one. For a small antiferromagnetic
 coupling (K>0), a completely
 asymmetric situation is obtained with one s=$\frac{1}{2}$ component strongly screened by the
 Kondo effect and the other one almost free to yield indirect magnetism, which
 shows finally a possible coexistence between a RKKY interaction and a 
 local Kondo effect, as observed in Uranium compounds such as $UPt_3$.
 
 \end{abstract}
 
\vskip 6cm

 
 \section{Introduction}

Kondo Cerium compounds have been extensively studied from both experimental and theoretical
point of view. In this case, the Kondo effect is well described by either the 
s-f exchange Hamiltonian with a $S^f=\frac{1}{2}$
spin screened by only one conduction electron channel\cite{un} or the so-called
Coqblin-Schrieffer Hamiltonian\cite{deux} when orbital degeneracy and spin-orbit
coupling are taken into account; in the two preceding cases, there is an equal
number of 4f and conduction electrons. The ground state of the regular Kondo effect
is a nonmagnetic singlet state in the case of a single impurity\cite{un} and the
low temperature properties are characterized by a Fermi-liquid behaviour. In the
case of heavy-fermion compounds, there is a strong competition between the Kondo
effect and the magnetic ordering, which yields either nonmagnetic or magnetically
ordered Cerium Kondo compounds\cite{tr,qua}.

On the other hand, some Uranium compounds, such as $UPt_3$, present also a heavy-fermion
behaviour and are also superconducting. $UPt_3$ has an outstanding behaviour, since
it undergoes a transition to an antiferromagnetic ordering with a tiny ordered magnetic 
moment of $0.02\pm0.01\mu_B$ below a Neel Temperature $T_N\sim 5 K$\cite{ci} and
becomes superconducting below $T_c\sim 0.5 K$\cite{si}. A heavy-fermion behaviour
characterized by a large electronic specific heat constant $\gamma\sim 0.4$ J/mole$K^2$\cite{se}
 and a $T^2$ term of the resistivity\cite{hu} is observed in $UPt_3$ at low temperatures. A
 third characteristic temperature $T_S=17.6 K$ given by the maximum of the magnetic
 susceptibility corresponds approximatively to the onset of spin fluctuations\cite{hu}. The
 heavy-fermion character decreases with
 pressure\cite{se,hu}, while the antiferromagnetic order disappears at roughly 5 kbar\cite{ne}.
 
 The real nature of the magnetic order in $UPt_3$ is still a controversial subject, because
 no small-moment antiferromagnetism has been observed in a recent $\mu$SR study
 of pure $UPt_3$\cite{di1}. Neutron-diffraction experiments\cite{di2} have been
 also recently carried out on single-cristalline samples of the heavy-fermion
 pseudobinary alloys $U(Pt_{1-x}Pd_x)_3$. At low Pd concentrations, x=0.002 and
 x=0.005, small-moment magnetic order is observed below 6K, just like in $UPt_3$. For
 large x values, a clear magnetic order exists , with a magnetic moment of O.35
 $\mu_B$ for x=0.02 and 0.62 $\mu_B$ for x=0.05 and with an increasing Neel
 Temperaure $T_N$=3.5K for x=0.02 and $T_N$=5.9K for x=0.05. These recent experiments
 on $UPt_3$-based alloys show that the magnetic order is a regular antiferromagnetic
 one for x$\geq$0.02, while the origin of the magnetic order is still controversial
 for $UPt_3$; the question arises to know if the magnetic moment of $UPt_3$ is
 a real long-range one or a so-called "short-range" with a very large correlation length.
 
 Other Uranium compounds namely $UBe_{13}$\cite{di3}, $URu_2Si_2$\cite{di4}, $UNi_2Al_3$\cite{di5,di6,di7} and $UPd_2Al_3$\cite{di5,di6,di7}
 present also at low temperatures a weak magnetic ordering with small magnetic
 moments, followed at still lower temperatures by a transition to a superconducting
 state. These Uranium compounds are characterized by a Neel temperature 
 ($T_N$=8.8K in $UBe_{13}\cite{di8}$, 17.5K in $URu_2Si_2$\cite{di4}, 4.6K in $UNi_2Al_3$\cite{di7} and 14K in
 $UPd_2Al_3$\cite{di7}) larger than the superconducting temperature (respectively 0.85K, 0.8K, 1K
 and 2K) and by large values of the electronic specific heat constant $\gamma$. The
 values of the magnetic moments are rather small, except in the case of $UPd_2Al_3$
 where an ordered magnetic moment of 0.85 $\mu_B$ has been deduced from an elastic
 neutron scattering study\cite{di9}. The origin of these small magnetic moments and the
 eventual similarity between Uranium and Cerium compounds as $CeCu_2Si_2$ have
 been discussed in many papers\cite{di10,di11,di12}. The exact nature of the magnetic ordering in these 
 Uranium compounds is not definitively established. However the existence of both
 a heavy-fermion character and a weak-magnetic ordering seems to be characteristic
 of Uranium compounds, while the question is more controversial in Cerium compounds
 such as $CeCu_2Si_2$, where the existence of a weak magnetic order has not been
 definitively established and any way depends on the sample composition\cite{di11}. According
 to Steglich et al,\cite{di11} recent experiments support the coexistence of two possible
 channels of so called "localized" and "itinerant" 5f states in Uranium compounds and these two 5f 
 subsystems appear to be only weakly coupled to each other in $UPd_2Al_3$ for example.

 Thus, the purpose of the present paper is to present an explanation for the coexistence
 of the heavy-fermion character and tiny ordered magnetic moments in Uranium
 compounds such as $UPt_3$. This explanation is based on the "underscreened Kondo model"
 which appears to be appropriate to describe the $5f^2$ configuration of Uranium
 atoms.
 
 The underscreened Kondo model corresponds to the case $2S>n$, where S is the localized
 spin and n the number of screening channels coming from conduction electrons\cite{on}. We will describe   
  here the simple case of the underscreened Kondo effect with S=1, n=1 but indeed it is
  certainly necessary to include the orbital degeneracy and spin-orbit effects
  to give a good description of compounds such as $UPt_3$.
  
The underscreened S=1 one-impurity Kondo Hamiltonian is given by:
 \begin{equation}
 \label{ttt}
 {\cal H}=\sum_{\vec{k}}{\epsilon}_{\vec{k}}\Psi^{\dag}_{\vec{k}}\Psi_{\vec{k}}+J\sum_{\vec{k},\vec{k'}}\Psi^{\dag}_{\vec{k}}\vec{\sigma}\Psi_{\vec{k'}}.\vec{S} 
\end{equation}
 where $\Psi^{\dag}_{\vec{k}}=(\Psi^{\dag}_{\vec{k},\uparrow},\Psi^{\dag}_{\vec{k},\downarrow}) $
  is a conduction electron spinor and S is a SU(2) 1-spin. The Hamiltonian (\ref{ttt}) has been 
  studied in the general context of the underscreened Kondo problem
 using renormalization methods\cite{cragg} and has been solved exactly by the Bethe Ansatz\cite{and} method and conformal
 theory arguments\cite{affl}.
 
 The ground state has a 2-fold degeneracy corresponding to an unquenched spin
 $\frac{1}{2} $, whose the residual coupling to the Fermi sea is ferromagnetic and scales to zero at low temperatures. The strong Fermi-liquid fixed point
   is {\bf stable}. The 
 low-energy electronic excitations are free-electron like and the many body
 interactions induced by the Kondo effect lead,  at low energy, to a simple phase shift which 
 is equal to $\delta=\frac{\pi}{2}$.

\section{General considerations on the two-impurity Kondo problem}

The two-impurity Kondo problem with a spin $s=\frac{1}{2}$ on each impurity, embedded
in a conduction electron band with only one n=1 channel, has been extensively
studied by many authors in the last ten years. A recent review of the main works
can be found in Ref.\cite{ph}. The two-impurity problem provides a simple model
to study the competition between the Kondo effect and the indirect Ruderman-Kittel-Kasuya-Yosida (RKKY)
interaction.

We would like to study here the two-impurity Kondo problem with a spin S=1 on each
impurity and with only one n=1 channel for conduction electrons. We consider
 two S=1 spins symmetrically
     located about the origin and interacting whith a Fermi gas. The total Hamiltonian is the sum of the three
     following terms:
     \begin{eqnarray}
     \label {aff}
      H_o&=&\int d^3 \vec{k}\hskip 0.2cm \epsilon(\vec{k})\Psi^{\alpha\dag}_{\vec{k}}\Psi_{\alpha\vec{k}}\\ \nonumber
      H_k&=&\int d^3 \vec{k_1}\int d^3 \vec{k_2} \hskip 0.2cm \Psi^{\alpha\dag}_{\vec{k_1}}\sigma^{\beta}_{\alpha}\Psi_{\beta {\vec{k_2}}}.[V(\vec{k_1})^*V(\vec{k_2})\vec{S_1}+V(-\vec{k_1})^*V(-\vec{k_2})\vec{S_2}],\\ \nonumber 
      H_i&=&K\vec{S_1}\vec{S_2}
       \end{eqnarray}
       where $ S_1$ and $S_2 $ are two S=1 impurities. $V(\vec{k})$
       is proportionnal to the Anderson model hybridization matrix element and we
       adopt here the particular choice of Ref.\cite{ph}. $V(\vec{k})$
       and K are considered as two independent parameters. The parameter K takes
       into account both, all the direct exchanges between $ S_1$ and $S_2 $
       and the RKKY interaction between two s=1/2 spins
      (one of $S_1$ and the second of $S_2$), defined by:
       \begin{equation}
       \label{rkky}
       K(R)=\frac{J^2}{E_F}\frac{\cos 2k_F R}{(2k_F R)^3}
       \end{equation}
       where, R is the distance between $ S_1$ and $S_2 $ and $E_F$ is the energy
       at the Fermi level.
       
       There are two stable obvious limits for this problem:
       \\
      -- when $K\longrightarrow +\infty$, the two S=1 spins tend to form a singlet of
       spin and, therefore, the electron gas is not affected by the presence of these two impurities. There 
       are no Kondo effect and a zero phase shift $\delta$ for
       the conduction band.
       \\
      -- when $K\longrightarrow -\infty$, on the contrary, the two impurities behave as an
       effective single S=2 impurity with n=2 channels of conduction
       electrons interacting with it. In this Kondo effect, only a S=1 spin of
       the effective impurity is screened; the remaining low-energy conduction electron
       degrees of freedom are decoupled from it, but yield 
       a $\delta=\frac{\pi}{2}$ phase shift in both channels. It corresponds to a local
       Fermi-liquid-fixed-point and, therefore, the many-body interactions lead to a simple phase shift at low energy.
   
      Thus, the purpose of the present paper is to study the S=1 two-impurity problem for all K values. The central
      question is, therefore, to know if the local Fermi-liquid
      description still holds for all K values at T=0 or similarly if the phase shift
      of the conduction electrons varies continuously with K at T=0.
       
       In the case of the two $s=\frac{1}{2}$ Kondo impurities, there must be, as
       a function of K, a phase transition, but the existence of a critical point 
       is still controversial, since for example numerical renormalization group calculations yield a critical
       point, while finite-temperature Monte Carlo (MC) calculations\cite{fye}
       do not show evidence for such a critical point. Thus, the question of an eventual
       phase transition has to be also discussed in our case of two
       S=1 spins.
       
       Thus, in the present section, we will present the main features of the two-impurity
       problem, which have been already developed for the $s=\frac{1}{2}$ case, in particular
       in the recent papers of Affleck et al.\cite{ph} and Sire et al.\cite{varma}. The 
       Hamiltonian (\ref{aff}) is transformed by using successively an one-dimension
       mapping and the classical bosonization technique, exactly as in the previous 
       $s=\frac{1}{2}$ case.
       
       In the next section III, we will describe our work on the specific S=1 two-impurity
       problem and we use successively the Jordan-Wigner transformation to refermionize
       the Hamiltonian and a specific mean field approximation to treat the problem. The
       different cases, especially K=0 and K>0 for an antiferromagnetic coupling, will
       be then discussed.
       
       \subsection{The one-dimensional mapping}
        
        We follow here the notations of the recent paper by Affleck et al\cite{ph}
        on the two $s=\frac{1}{2}$ impurity case and we just recall the main points 
        for our study of the two S=1 impurities.   
        
        As usual, we consider a $\delta$ function interaction in (\ref{aff}), with the impurities
        at $\pm\frac{\vec{R}}{2}$, so that:
        \begin{equation}
        \label{ddd}
         V(\vec{k})=V_o e^{i\vec{k}\frac{\vec{R}}{2}}
         \end{equation}
       
        For certain choices for the dispersion relation $\epsilon (\vec{k})$ and 
        matrix element $V(\vec{k})$, the Hamiltonian (\ref{aff}) has a particle-hole (PH) symmetry. Invariance of $H_o$ 
        under the  PH  symmetry requires: $\epsilon (\vec{k})=-\epsilon (\vec{k'})$, where
        $\vec{k}$ and  $\vec{k'}$ are changed to each other by the PH symmetry.  
        \\
        Invariance of $H_k$ under this PH symmetry requires:
        \begin{equation}
        \label{ref1}
        V(\vec{k'})=V(\vec{k})^* \qquad \text{and}\qquad V(-\vec{k'})=V(-\vec{k})^*e^{i\eta}
        \end{equation}
         where $\eta $ is just a phase independent of k\cite{ph}.
        
        To apply the bosonization technique to this problem, one first shows that ${\cal H}$ can
        be reduced exactly to an one-dimensional Hamiltonian. For that, one makes a projection
        on iso-energy surfaces in $\vec{k}$ space; two $\vec{k}$ are only retained by the Kondo effect
        and one can define the two following fields: 
        \begin{equation}
        \label{ref2}
        \Psi_{\pm,E}=\int d^3 \vec{k}\hskip 0.2cm\delta(\epsilon(\vec{k})-E)V({\pm}\vec{k})\Psi_{\vec{k}}
        \end{equation}
        Hence, odd and even combinations of these two fields are defined:
        \begin{equation}
        \label{ref3}
        \Psi_{e,E}=\frac{\Psi_{+,E}+\Psi_{-,E}}{N_e(E)}, \qquad 
         \Psi_{o,E}=\frac{\Psi_{+,E}-\Psi_{-,E}}{N_o(E)}
         \end{equation}
         where
          \begin{equation}
          \label{st}
          N_{e,o}(E)=\int d^3 \vec{k}\ \delta(\epsilon(\vec{k})-E).|V(\vec{k}){\pm}V(-\vec{k})|^2
        \end{equation}
         to satisfy the anticommutation rules: $\{\Psi_{E},\Psi_{E}^{\dag}\}_+=\delta(E-E')$
         
         Only these two fields appear in ${\cal H}$ and we can rewrite:
         \begin{eqnarray}
         \label{ggg}
         H_o&=&\int dE\ E[\Psi_{e,E}^{\dag}\Psi_{e,E}+\Psi_{o,E}^{\dag}\Psi_{o,E}],\\ \nonumber 
         H_k&=&\int dEdE'\ [N_e(E)N_e(E')\Psi_{e,E}^{\dag}\vec{\sigma}\Psi_{e,E'} + N_o(E)N_o(E')\Psi_{o,E}^{\dag}\vec{\sigma}\Psi_{o,E'}].(\vec{S_1}+\vec{S_2}) \\ \nonumber
          &+&N_e(E)N_o(E').(\Psi_{e,E}^{\dag}\vec{\sigma}\Psi_{o,E'}+\Psi_{o,E}^{\dag}\vec{\sigma}\Psi_{e,E'}).(\vec{S_1}-\vec{S_2})
         \end{eqnarray}
     
         Indeed, the one-dimensional problem has also a PH symmetry, deduced from the \hbox{three-dimensional}
         PH one. The problem of the particle-hole symmetry has been previously
         studied for the two-impurity $s=\frac{1}{2}$ case \cite{ph,ph2}, because
         in some special cases, one can develop  some qualitative arguments for the variation with
         K of the phase shift of the conduction electrons and here, therefore, some insight
         on a possibility of a phase transition at a given K value.
         
         The transformation of the 
         fields $\Psi_{e,E} $ and $\Psi_{o,E}$ can be deduced from:
         \begin{equation}
         \Psi_{+,E}\longrightarrow\Psi_{+,-E}^{\dag},\qquad\Psi_{-,E}\longrightarrow\Psi_{-,-E}^{\dag}e^{{i\eta}}
         \end{equation} obtained with the initial PH transformation.
         \\ 
         In our case, we follow the method of Ref.\cite{ph} and we can
          select two particular values of $\eta$, i.e. ($\eta$=0 and $\eta$=$\pi$), which give  arguments for
         two different physical behaviours. 
                
         For $\eta$=0, using the preceding PH one-dimensional transformation, one
         obtains:
         \begin{equation}
         \label{nono}
         N_e(-E)=N_e(E),\qquad N_o(-E)=N_o(E),\qquad\Psi_{e,E}\longrightarrow\Psi_{e,-E}^{\dag},\qquad \Psi_{o,E}\longrightarrow\Psi_{o,-E}^{\dag}
          \end{equation}
          \\
          and for $\eta=\pi$, it results:
          \begin{equation}
          \label{neuf}
          N_e(-E)=N_o(E),\qquad\Psi_{e,E}\longrightarrow\Psi_{o,-E}^{\dag},\qquad \Psi_{o,E}\longrightarrow\Psi_{e,-E}^{\dag}
          \end{equation} 
         
         The case $\eta=\pi$ is of particular interest. According to Ref.\cite{ph}, the
         phase shifts $\delta_e$ and $\delta_o$ for the two fields given by
         (\ref{ref3}) can take arbitrary values with the easily satisfied condition $\delta_e=-\delta_o$. Using
         the PH symmetry, it results only that a transition is not necessary in this case. The case $\eta=0$, imposes 
          $\delta_e=\delta_o=0$ or $\delta_e=\delta_o=\frac{\pi}{2}$ and consequently
         a  transition in the phase diagram. Finally, no universal behaviour can be predicted with a simple phase shift
         analysis.  
         
         Thus, we continue the calculation and write $ H_k$ given by (\ref{ggg})
          in a more suitable form, around E=0:
         \begin{eqnarray}
         \label{tf}
         H_k&=&\frac{1}{2}\int dEdE'\ J_+(\Psi^{\dag}_{1,E}\vec{\sigma}\Psi_{1,E'}+\Psi^{\dag}_{2,E}\vec{\sigma}\Psi_{2,E'}).(\vec{S_1}+\vec{S_2})\\ \nonumber
            &+&J_m(\Psi^{\dag}_{1,E}\vec{\sigma}\Psi_{1,E'}-\Psi^{\dag}_{2,E}\vec{\sigma}\Psi_{2,E'}).(\vec{S_1}-\vec{S_2})\\ \nonumber
            &+&J_-(\Psi^{\dag}_{1,E}\vec{\sigma}\Psi_{2,E'}+\Psi^{\dag}_{2,E}\vec{\sigma}\Psi_{1,E'}).(\vec{S_1}+\vec{S_2})
         \end {eqnarray}
         with the orthonormal basis:
	 \begin{equation}
	 \Psi_{1,2}=\frac{(\Psi_e\pm\Psi_o)}{\sqrt{2}}\qquad
          \end{equation}
          and the couplings:
          \begin{equation}
          \label{qn}
          J_{\pm}=\frac{(J_e{\pm}J_o)}{\sqrt{2}} ,\qquad
          J_e=2N_e(0)^2,\qquad
         J_o=2N_o(0)^2,\qquad J_m=2N_o(0)N_e(0)
        \end{equation}
        When the asymmetry between the odd and even channels is not relevant, as
        for example in the case $\eta=\pi$, we get the following simplification:
         \begin{equation}
          \label{bobo}
         N_e(0)=N_o(0),\qquad J_m=J_+,\qquad J_-=0
         \end{equation}
        which will be used in the following.  
         
         We can notice that, whatever the maintained particle-hole symmetry is, the charges of the 1 and 2 species of fermions
         are separetely conserved and we have two commuting sets of isospin generators; in fact, there is
         an exact $O(4)=SU(2)_I*SU(2)_S$ symmetry on each channel:
         \begin{eqnarray}
          I^z_1&=& \frac{1}{2} \int dE\ \Psi^{\dag\alpha}_{1,E}\Psi_{\alpha 1,E}\\ \nonumber
          I^-_1&=& \int dE\  \Psi_{\uparrow 1,E}\Psi_{\downarrow 1,-E} 
          \end{eqnarray}
          similarly for $\vec{I}_2$.
          \\
          We just analyze the situation with
          an abelian symmetry (for the charge and spin degrees of freedom) and consequently, we have to break {\bf explicitly} the O(4) one. We
          expect that the low-energy physics remains the same because a representation of the  SU(2), k=1
          level algebra, with  a central charge c=1 can be satisfied by a representation of  free
          bosons.
          
       If now we try to calculate the preceeding values by taking
        the particular choice $\epsilon(\vec{k})=v_F(k-k_F)$ and by making
        the integration in the Eq.(\ref{ref2}), we obtain:
       \begin{equation}
       N_{e,o}(k)=V_o\sqrt{1\pm\frac{\sin\ (kR)}{kR}} 
       \end{equation} 
       \\
       From the equalities (\ref{qn}), the couplings, $J_m,\ J_+,\ J_-$ can be now easily evaluated:
         \begin{equation}
         \label{www}
         J_+=\pi\rho_F V_o^2,\qquad J_-=\pi\rho_F V_o^2*\frac{\sin\ (k_F R)}{k_F R},\qquad J_m=\pi\rho_F V_o^2 \sqrt{1-(\frac{\sin\ (k_F R)}{k_F R})^2}
         \end{equation}
         where $\rho_F$ is the density of states of the conduction electrons 
         per spin at the Fermi level.
         \\
        Thus, $J_-=0$ 
         is equivalent to the equality $k_F R=n\pi$, where n is an integer; at half-filling
         this constraint is realized for $k_F=\frac{\pi}{2c}$ and consequently for an even impurity
         distance $R=2nc$. 
         \\
         In these conditions, by using the eq. (\ref{rkky}) we
         deduce that a conventional RKKY interaction could only exist in the case
         of a parameter K>0 (since $2k_F R=2n\pi$).
         
         Now, we make some comments about the  feasible physical interpretation of $\eta$ of the  Ref.\cite{ph}. Indeed, if we take the definition (\ref{ddd})
         of $V(\vec{k})$ and the second relation (\ref{ref1}), we immediately obtain:
         \begin{equation}
         \label{eee}
         \eta=\frac{\vec{k_o}.\vec{R}}{2}
         \end{equation}
         where $\vec{k_o}=\vec{k'}-\vec{k}$ and not $\eta={\vec{k_o}}.{\vec{R}}$ already mentionned
         by Affleck et al. in Ref.\cite{ph}.
         
         In this context, the two values of $\eta$ correspond either to $\vec{k_o}$=0 or to
         the nesting vector $\vec{k_o}=(\frac{\pi}{c},\frac{\pi}{c},\frac{\pi}{c})$. 
         With the condition R=2nc and the Eq. (\ref{eee}), we find that  $\eta=\pi*n$, which
         yields $\eta$=0 or $\eta=\pi$ depending on the parity of n. We see that there is no universal
         behaviour and we cannot use this physical argument to conclude on the possibility
         of a critical point, in contrast to Ref.\cite{ph}; furthermore in the following, we
         will check that there is no critical point or even no phase transition in our
         S=1 case.
          
          By Fourier transform, we immediately
          obtain ${\cal H}$, in terms of two one-dimensional electronic channels a and b:
          \begin{equation}
          \label{seize}
         {\cal H}=H_o+K\vec{S_1}\vec{S_2}+J_+.(\vec{S_1}+\vec{S_2})[a^{\dag}\frac{\vec{\sigma}}{2}a+b^{\dag}\frac{\vec{\sigma}}{2}b]_{x=0}
                 +J_m.(\vec{S_1}-\vec{S_2})[a^{\dag}\frac{\vec{\sigma}}{2}a-b^{\dag}\frac{\vec{\sigma}}{2}b]_{x=0}
          \end{equation}
          \\
          with 
          \begin{equation}
          \label{we}
         a(x)=\int dk\  e^{ikx}\Psi_1(k),\qquad b(x)=\int dk\ e^{ikx}\Psi_2(k)
         \end{equation} 
          \\
          \text{and}
          \begin{equation}
          \label{wq}
         J_m=J_+=J=\pi\rho_F V_o^2, \qquad H_o=(i v_F)\sum_{\sigma}\int_{-\infty}^ {+\infty} dx\ (a^{\dag}_{\sigma}(x)\partial_x a_{\sigma}(x)+b^{\dag}_{\sigma}(x)\partial_x b_{\sigma}(x))  
         \end{equation}
         
          Within this model with $J_m=J_+$, no indirect magnetic interaction, i.e. via the conduction band, is
          generated up to the second order in
         perturbation between two half-spins, respectively of $S_1$ and $S_2$. Hence, we can assume that the Heisenberg interaction $K\vec{S_1}\vec{S_2}$
         takes into account both the indirect RKKY interaction and the direct one between the two
        S=1 spins. 
              
         \subsection{The bosonization}
   
          Thus, in the following, we start from the form (\ref{seize}) of the
          Hamiltonian for two S=1 spins and, as previously done for the case of
          two $s=\frac{1}{2}$ spins, we use the bosonization technique
          by taking the standard 1-dimensional relations
          between Bose and Fermi fields\cite{eme}:
          \begin{eqnarray}
          \Psi(x)&=&\frac{1}{2{\pi} c}\exp{i\Phi_{\psi}(x)}\\ \nonumber
          \psi&=&a_\uparrow, \hskip 0.1cm a_\downarrow, \hskip 0.1cm b_\uparrow,\hskip 0.1cm b_\downarrow\\ \nonumber
          \Phi_\psi&=&-\sqrt{\pi}(\phi_\psi+\int_{-\infty}^x\pi_\psi(x')dx')
          \end{eqnarray}
         $ \phi_\psi$ and $ \pi_\psi$ are respectively a bosonic field and its conjugate field; as usual, the lattice 
         spacing c is taking as tending to zero.
         In ${\cal H}$, the two electronic channels a and b are independent, then there is no need
         to introduce any phase factor in the $\Psi$ field definition to take
         care of the anticommutation rules between the two different "species" of fermions\cite{gan}.
         Then, we just redefine four new bosonic fields that we call respectively charge, spin, spin-channel
          and charge-channel fields, obtained from the preceding ones by a linear canonical
          transformation:
          \begin{equation}
          \label{pi}
          \Phi_c=\frac{1}{2}(\Phi_{c,a}+\Phi_{c,b}),\qquad
           \Phi_s=\frac{1}{2}(\Phi_{s,a}+\Phi_{s,b}),\\
           \end{equation}
           \begin{equation}
           \label{pi2}
            \Phi_{sf}=\frac{1}{2}(\Phi_{s,a}-\Phi_{s,b}),\qquad
           \Phi_{cc}=\frac{1}{2}(\Phi_{c,a}-\Phi_{c,b})
          \end{equation}
          where $\Phi_{c,i}$ and $\Phi_{s,i}$ are the charge and the spin fields for the i=a, b channels.
          The degrees of charge are frozen, thus it is clear that $\Phi_c$ and $\Phi_{cc}$ are not coupled to the impurities and we can omit them.
          
          Thus, it comes:
          \begin{eqnarray}
          \label{ft}
          H_o&=&\frac{v_F}{2} \sum_{\lambda=s,sf}\int_{-\infty}^{+\infty} dx\ \{\Pi_{\lambda}^2+[\partial_x\phi_{\lambda}]^2\}\\ \nonumber
          H_k&=&\frac{J_{z,+}}{\pi}(S_1^z+S_2^z)\nabla\Phi_s(0)+\frac{J_{m,z}}{\pi}(S_1^z-S_2^z)\nabla\Phi_{sf}(0)\\ \nonumber
            &+& \frac{J}{{\pi}c}\cos\Phi{sf}(0)\{\cos\Phi{s}(0)(S_1^x+S_2^x)-\sin\Phi{s}(0)(S_1^y+S_2^y)\}\\ \nonumber
           &-&  \frac{J}{{\pi}c}\sin\Phi{sf}(0)\{\sin\Phi{s}(0)(S_1^x-S_2^x)+\cos\Phi{s}(0)(S_1^y-S_2^y)\}
          \end{eqnarray}
          \\
        The Kondo couplings $J_{z,+}$ and $J_{m,z}$ can take different values in the (x,y) plane and
          along the z axis. We can then perform a rotation along the quantization axis
           to eliminate $\Phi{s}$. This type of procedure which originated from Ref.\cite{emery} was, for instance, used
            in 
           the study of the two-channel one-impurity Kondo problem\cite{kiv,giam,geo}. This can be achieved
           by considering  the canonical transformation, in 
           the unit sphere, $U=\exp(-i(S_1^z+S_2^z))\Phi_s(0)$. 
       The effect of the rotation is to replace the trigonometric functions of $\Phi_s(0)$ in Eq. (\ref{ft}) by their
       values at zero argument.
     
      We obtain therefore:
          \begin{eqnarray}
           \label{tre}  
            H_k&=&\frac{J_{z,+}-\rho_F^{-1}}{\pi}(S_1^z+S_2^z)\nabla\Phi_s(0)+\ \frac{J_{m,z}}{\pi}(S_1^z-S_2^z)\nabla\Phi_{sf}(0)\\ \nonumber
          &+& \frac{J}{{\pi}c}\{\cos\Phi{sf}(0)(S_1^x+S_2^x)-\sin\Phi{sf}(0).(S_1^y-S_2^y)\}
           \end{eqnarray}
           
         where  $\rho_F $ is the density of states at the Fermi level for the conduction electrons $\Psi_s=e^{i\Phi_s(x)}$.
            The $H_i$ coupling is not really affected by the transformation:
           \\
           \begin{equation}
           H_i=K_zS_1^zS_2^z+KS_1^{+}S_2^{-} .
           \end{equation}\par
           The \hskip 0.07cm canonical \hskip 0.07cm transformation \hskip 0.05cm also \hskip 0.05cm generates \hskip 0.05cm a \hskip 0.05cm positive \hskip 0.05cmconstant \hskip 0.05cm term \hskip 0.05cm proportional \hskip 0.05cm to \par{\noindent
           $(S_1^z+S_2^z)^2=2S_1^zS_2^z+\frac{1}{2}$, which can be reabsorbed
           in the  $K_z$ term and a negative coupling $ J_{z,+}(S_1^z+S_2^z)^2$\cite{gan}.
           
          Now, we have to fix the $J_{z,+}$ and $J_{m,z}$
           couplings. The $J_{z,+}$ one can be integrated out using a path integral formalism; it
            only renormalizes the RKKY interaction \cite{gan}. This can be realized
            directly in tuning K and $K_z$. Thus, in the following, we set $J_{z,+}=\rho_F^{-1}$. 
         When K=0 we have chosen the particular case where the charges of the a and b electronic channels are separately conserved
         and for that we could not keep $J_{m,z}$ as a tunable parameter; from 
         the Hamiltonian (\ref{tre}), in using Eqs. (\ref{pi}), (\ref{pi2})
         we have to take $J_{m,z}\rightarrow 0$. By analogy with the Toulouse
         limit, first discovered in
         the ordinary Kondo problem\cite{tou}, we expect that this model is solvable
         at the particular point:  $J_{z,+}=\rho_F^{-1}$ and $J_{m,z}$=0. Now, we look at the Kondo problem in the transverse direction
         as keeping $J=J_m$, but allow $J_{m,z}\not=J_{z,+}=\rho_F^{-1}$:
          \begin{equation}
          \label{pi3}
          {\cal H}=H_o + \frac{J}{2{\pi}c}\{S_1^+e^{i\Phi_{sf}(0)}+e^{i\Phi_{sf}(0)}S_2^{-}\}+H_i
          \end{equation}

          \section{the study of the two-impurity S=1 underscreened Kondo problem}

           In the preceding section, we have presented the general formalism appropriate for the two-impurity
           Kondo problem and we have finally obtained the form (\ref{pi3}) of the total Hamiltonian, which is 
          valid for any value of the spin. Then, we study the specific case of two S=1 spins and for that we decompose the
two S=1 spins, $S_1$ and $S_2$, into two 1/2-spins, as follow:
       
           \begin{eqnarray}
           \label{pi4}
           \vec{S_1}&=&\vec{\tau_1}+\vec{\tau_2}\\ \nonumber
           \vec{S_2}&=&\vec{\tau_3}+\vec{\tau_4}
           \end{eqnarray}
           where, $\{\vec{\tau_i}\}_{(1,2,3,4)}$ are half SU(2) spins, which satisfy:
           \begin{eqnarray}
          \{\tau_i^{+},\tau_j^{-}\}_-&=&0\\ \nonumber
          \{\tau_i^{+},\tau_i^-\}_+&=&1,\text{\hskip 0.5cm for i=1, 2, 3, 4 and j=1, 2, 3, 4}
          \end{eqnarray}
          
          Indeed, we could not enlarge the total Hilbert space of the problem;  so we add the constraint
          that $\vec{\tau_1},\vec{\tau_2}$
          and $\vec{\tau_3},\vec{\tau_4}$ are strongly ferromagnetically coupled through an infinite -$M_z$ ($M_z$>0) coupling.
           Then, we will solve the Hamiltonian given by (\ref{pi3}) with the transformations (\ref{pi4}). To do it, we
           refermionize the Hamiltonian (\ref{pi3}) with spinless fermions by use of the Jordan-Wigner transformation and
           then we use a mean-field approximation which keeps terms containing at most four operators.

           \subsection{Refermionization}
        
           In the following, we use the conduction electron operator:
           \begin{equation}
           \label{mm}
           \Psi=\frac{1}{2\pi c}\exp(i\Phi_{sf}(0))
           \end{equation}

           To refermionize this problem of four sets of Pauli matrices, we use the
           Jordan-Wigner transformation \cite{jor} for four spins:
           \begin{eqnarray}
           \label{mmm}
            \tau_1^{+}&=& d^{\dag}_1\\ \nonumber
            \tau_2^{+}&=&d^{\dag}_2.\exp(i{\pi}n_1)\\ \nonumber 
            \tau_3^{+}&=&d^{\dag}_3.\exp(i{\pi}(n_1+n_2))\\ \nonumber 
            \tau_4^{+}&=&d^{\dag}_4.\exp(i{\pi}(n_1+n_2+n_3))\\ \nonumber  
            \tau^z_i&=&{d^{\dag}_i}d_i-\frac{1}{2},\text{\hskip 0.2cm with $n_i=0,\ 1$ for i=$1,\ 2,\ 3,\ 4$}
            \end{eqnarray}
             
           Then, we develop $\cal H$ in power of $(n_i.n_j)$, with i=$1,\ 2,\ 3,\ 4$ and j=$1,\ 2,\ 3,\ 4$, in using the shrewd identity: 
           \begin{equation}
           \label{mmmm}
           \exp(i{\pi}n) = 1-2n\qquad \text{for n=0 or 1}
           \end{equation}
           
           Thus, applying the different transformations (\ref{mm}),(\ref{mmm}),(\ref{mmmm}) on the reduced
           Hamiltonian (\ref{pi3}) yields many terms containing products of operators $\Psi$ and $d_i$; in
           particular, we get several terms containing more than 4 spinless fermion operators, such as for
           example $d^{\dag}_1 d_4 n_2 n_3$ (obtained from $H_i$) or $\Psi d_4 n_2 n_3$ (obtained from
           $H_k$). In order to solve the problem we use here a \underline{special mean-field approximation}, which
           consists in firstly keeping only terms containing at most four operators and then making averages on terms with two operators. In
           fact, as we will see in the following, we will use a mean-field approximation
           which linearizes the terms in the Hamiltonian and keeps only terms
           which are bilinear in the spinless fermion operator d. This approximation had already been used in
           Ref.\cite{varma} for the $s=\frac{1}{2}$ two-impurity problem. We have to remember the two following
           points induced by the transform (\ref{pi4}): $\tau_1$, $\tau_2$ interact with the same conduction electron and $\tau_3$, $\tau_4$ with another
           one, while $ \langle\vec{\tau_1}\vec{\tau_3}\rangle$=$ \langle\vec{\tau_2}\vec{\tau_4}\rangle$  and 
         $ \langle\vec{\tau_1}\vec{\tau_4}\rangle$=$ \langle\vec{\tau_2}\vec{\tau_3}\rangle$, but 
           not necessairely $ \langle\vec{\tau_1}\vec{\tau_3}\rangle$=$ \langle\vec{\tau_1}\vec{\tau_4}\rangle$.
          
          Then, using the preceding approximations, the total Hamiltonian can be written  in terms of
          fermionic spinless operators:
           \begin {eqnarray}
           \label{trf}
           {\cal H}&=&H_k+H_{int}+H_o,\text{\hskip 0.1cm with\hskip 0.1 cm}\qquad \\ \nonumber
           H_o&=&iv_F\int_{-\infty}^{+\infty} dx\Psi^{\dag}(x)\partial_x\Psi(x)\\ \nonumber
           H_{k}&=&H_a+H_+ \\ \nonumber
           H_{int}&=&H_{i,\|}+H_{i,\bot}+H_{ferro}
           \end{eqnarray}
           
           $H_a $ describes the Kondo problem when the two S=1 spins are not coupled, $H_+$ 
           brings a new Kondo contribution coming from the K interaction 
           and $ H_{ferro}$ is added here to take into account the decomposition of the S=1 spins
           (with the assumption $M_z\rightarrow \infty$):
           \begin{eqnarray}
           \label{trg}
           H_a&=& J\Psi(d^{\dag}_1+d^{\dag}_2)-J\Psi{d^{\dag}_1}n_2-J\Psi{d^{\dag}_2}n_1\\ \nonumber
           &-&J\Psi(d_3+d_4)+ J\Psi d_3 n_4 + J\Psi d_4 n_3\\ \nonumber
          H_+&=&J\Psi(d_3+d_4)\{n_1+n_2\}-J\Psi(d^{\dag}_1+d^{\dag}_2)\{n_3+n_4\}\\ \nonumber
          H_{i,\bot}&=&\frac{K}{2}\{d^{\dag}_1d_3(1-2n_2-2n_4)+d^{\dag}_1d_4(1-2n_2-2n_3)\\ \nonumber
                     &+&d^{\dag}_2d_3(1-2n_4-2n_1)+ d^{\dag}_2d_4(1-2n_3-2n_1)+hc\}\\ \nonumber
          H_{i,\|}&=&K_z\{d_1{\dag}d_1+d_2{\dag}d_2-1\}\{d_3{\dag}d_3+d_4{\dag}d_4-1\}\\ \nonumber 
          H_{ferro}&=&-M_z(d_1^{\dag}d_1-\frac{1}{2})(d_2^{\dag}d_2-\frac{1}{2})-M_z(d_3^{\dag}d_3-\frac{1}{2})(d_4^{\dag}d_4-\frac{1}{2})
          \end{eqnarray}
    
	     \subsection{The case $K=0$}
             As previously explained, our presently studied case $J_m=J_+$ corresponds
             to a situation where the indirect and direct interactions between
             $\vec{S}_1$ and $\vec{S}_2$ are yielded only by the additive term
             K$\vec{S}_1\vec{S}_2$. Thus, the case K=0 corresponds to two initial
             S=1 spins which are decoupled from each other and the physics of this
             problem is similar to that of the one S=1, n=1 Kondo impurity. It results 
             that the term $H_+$ of the Hamiltonian must have no effect and that
             ${\cal H}$ can be divided into two independent underscreened Kondo
             problems:
             \hskip 4cm
             \begin{equation}
              H_{H_i=0}=H_{1,2}+H_{3,4}+H_o
             \end{equation}
	     with
             \begin{eqnarray}
             \label{v}
             H_{1,2}&=&J\Psi(d^{\dag}_1+d^{\dag}_2)-J\Psi{d^{\dag}_1}n_2-J\Psi{d^{\dag}_2}n_1\\ \nonumber
             &-&M_z(d_1^{\dag}d_1-\frac{1}{2})(d_2^{\dag}d_2-\frac{1}{2})
             \end{eqnarray}
             and
             \begin{eqnarray}
             \label{vv}
             H_{3,4}&=&-J\Psi(d_3+d_4)+J\Psi{d_3}n_4+J\Psi{d_4}n_3\\ \nonumber
             &-&M_z(d_3^{\dag}d_3-\frac{1}{2})(d_4^{\dag}d_4-\frac{1}{2})
             \end{eqnarray}
     We study this case K=0, in order to fix the theoretical notations for the
     following studies. 
     
     Then, in our present case, we can easily derive the following equalities for
     the average values:
     \begin{eqnarray}
     \label{kamel}
     \langle d^{\dag}_1d_2\rangle&=& \hskip 0.25cm\langle d^{\dag}_3d_4\rangle,\qquad\langle d^{\dag}_1d_2^{\dag}\rangle=\hskip 0.25cm\langle d^{\dag}_3d_4^{\dag}\rangle\\ \nonumber
     \langle\Psi d^{\dag}_1\rangle &=& -\langle\Psi d_3\rangle,\qquad\langle\Psi d^{\dag}_2\rangle =-\langle\Psi d_4\rangle \\  \nonumber
     \langle\Psi d_1\rangle&=&-\langle\Psi d^{\dag}_3\rangle,\qquad\langle\Psi d^{\dag}_4\rangle=-\langle\Psi d_2\rangle
     \end{eqnarray}
     We describe here the S=1 spins by adding two $s=\frac{1}{2}$ spins ferromagnetically
     aligned, according to the last terms of (\ref{v}) and (\ref{vv}) with $M_z$
     tending to $+\infty$. As in ref. \cite{varma}, the last term of (\ref{v})
     can be decoupled in the mean-field approximation into:
            
     $$ \frac{-M_z}{2}\{\langle n_2-\frac{1}{2}\rangle(n_1-\frac{1}{2})+\langle n_1-\frac{1}{2}\rangle(n_2-\frac{1}{2})\}
     + \frac{M_z}{2}\langle d_1^{\dag}d_2\rangle\{d_1^{\dag}d_2+d_2^{\dag}d_1\}
      -\frac{M_z}{2}\langle d_1^{\dag}d_2^{\dag}\rangle\{d_1^{\dag}d_2^{\dag}+d_2d_1\}$$
     
     The first term does not contribute due to the effective particle-hole (PH) symmetry. However, our
     present case of two S=1 spins is clearly original and we have to examine
     the solution occuring for a very strong ferromagnetic coupling $M_z\rightarrow+
     \infty$. If we consider the energy of the system, it is necessary to stabilize
     it to take both $\langle d_1^{\dag}d_2\rangle$ tending to 0 for $M_z\rightarrow+
     \infty$ and $\langle d_1^{\dag}d_2^{\dag}\rangle$ tending to its maximum value, which
     must be equal to $\langle d_1^{\dag}d_2^{\dag}\rangle=1$. 
     
     It results that the first three terms of $H_{1,2}$ given by (\ref{v}), treated
     within the preceding mean field approximation, become equal to
     $J\Psi[d_1^{\dag}+d_2^{\dag}+(d_1-d_2)\langle d_1^{\dag}d_2^{\dag}\rangle]$ and
     it results a new important contribution $J\Psi(d_1-d_2)$ for $\langle d_1^{\dag}d_2^{\dag}\rangle=1$,
     in addition to the first term $J\Psi(d_1^{\dag}+d_2^{\dag})$.
     
     For the physical limit $M_z\rightarrow+\infty$ corresponding to a S=1 spin, we
     can rewrite $H_{1,2}$ in the following form:
     \begin{equation}
     H_{1,2}=J\Psi(d_1^{\dag}+d_2^{\dag}+d_1-d_2)+h(d_1^{\dag}d_2^{\dag}+d_2d_1)+H_o
     \end{equation}
     where h is determined by the following self-consistent equations: \begin{equation}
     h=- \frac{M_z}{2}+J\langle\Psi d_1\rangle \rightarrow-\infty
     \end{equation}
     and 
     \begin{equation}
     \label{kiki}
     \langle\Psi d_1^{\dag}\rangle=\langle\Psi d_2^{\dag}\rangle=
     \langle\Psi d_1\rangle=-\langle\Psi d_2\rangle
     \end{equation}
     
     We can deduce that, due to the strong pairing mechanism between $d_1$ and $d_2$, only one degree of freedom is coupled to the conduction
     band:
     half a degree of freedom for the $1^{st}$ spinless fermion $\frac{(d_1^{\dag}+d_1)}{2}$
     and half a degree of freedom for the $2^{nd}$ spinless fermion $\frac{(d_2^{\dag}-d_2)}{2}$. 
     
     Then, to make the Kondo
     problem more explicit, we redefine 
     two new spinless fermions d and D by the simple linear transformation:
     \begin{equation}
     \label {fg1}
     d^{\dag}=(a_1+ib_2),\qquad d=(a_1-ib_2),\qquad D^{\dag}=(a_2+ib_1),\qquad D=(a_2-ib_1)
     \end{equation}
     with
     \begin{equation}
     \label{fg2}
     a_1=\frac{(d_1^{\dag}+d_1)}{2},\qquad a_2=\frac{(d_2^{\dag}+d_2)}
      {2},\qquad ib_1=\frac{(d_1^{\dag}-d_1)}{2},\qquad ib_2=\frac{(d_2^{\dag}-d_2)}{2}
     \end{equation}
     
     One can easily check that the different operators satisfy the good anticommutation rules.
    \\
    Only the d fermion is resonant and is coupled to the conduction band through
    a coupling $J^*=2J$. Consequently, the d and D fermions are not coupled anymore
    , the h coupling just shifts the resonant d-level at the
    Fermi energy $E_d=E_F=-h$ and makes the D-level lying at the energy $E_D=h$. As 
    usual, we redefine the Fermi energy $E_F=0$ and consequently $E_D=2*h$. Thus, the
    Hamiltonian $H_{1,2}$ can be written as:
    \begin{equation}
    H_{1,2}=J^*\Psi d^{\dag}+ E_D.D^{\dag}D
    \end{equation}
     \vskip 0.3cm
  
    Then, for
    the fermions $d_3$ and $d_4$, we propose the same relations:
    \begin{equation}
   \label{fg3}
    e^{\dag}=(a_3+ib_4),\qquad e=(a_3-ib_4),\qquad E^{\dag}=(a_4+ib_3)
      ,\qquad E=(a_4-ib_3)
      \end{equation}
      with
     \begin{equation} 
     \label{fg4}
     a_3=\frac{(d_3^{\dag}+d_3)}{2},\qquad
      a_4=\frac{(d_4^{\dag}+d_4)}{2},\qquad ib_3=\frac{(d_3^{\dag}-d_3)}{2},\qquad ib_4=\frac{(d_4^{\dag}-d_4)}{2}
    \end{equation}
     \\
     Finally, for $K=0$, the total Hamiltonian ${\cal H}$, can be written as two usual "not-coupled" resonant levels \cite{no3}:
     \begin{equation}
     {\cal H}_{K=O}=J^*\Psi d^{\dag}+J^*\Psi^{\dag}e^{\dag}+H_o^*
     \end{equation}
    with:
    \begin{equation}
    H_o^*=H_o+E_D.(D^{\dag}D + E^{\dag}E)
    \end{equation}
    
     It is well-known that this model is isomorphic to the usual Kondo effect
     at a certain particular point namely the Toulouse limit\cite{tou}:
     \begin{eqnarray}
     \label{bibi}
     {\cal H}_{K=O}&=&J^*s_1^{\dag}(a^{\dag}_{\downarrow}a_{\uparrow})+J^*s_3^{\dag}(b^{\dag}_{\downarrow}b_{\uparrow})\\ \nonumber
     &+&G_z s_1^z(a^{\dag}_{\uparrow}a_{\uparrow}-a^{\dag}_{\downarrow}a_{\downarrow})+G_z s_3^z(b^{\dag}_{\uparrow}b_{\uparrow}- b^{\dag}_{\downarrow}b_{\downarrow})+H_o
      \end{eqnarray}\\
      with
      \begin{equation}
      G_z=0=J_z^*-\frac{2}{\rho_F}
      \end{equation}
      where
      \begin{equation} 
       \vec{s_1}\hskip 0.4cm \text {and}\hskip 0.4cm \vec{s_2}\in \vec{S_1}\qquad \text {with}\qquad |\vec{s_1}|=|\vec{s_2}|=\frac{1}{2}\\ \nonumber
       \end{equation}
       and:
       \begin{equation} 
        \vec{s_3}\hskip 0.4cm\text {and}\hskip 0.4cm \vec{s_4}\in \vec{S_2}\qquad \text {with}\qquad |\vec{s_3}|=|\vec{s_4}|=\frac{1}{2}
        \end{equation}  
      
      The model (\ref{bibi}) describes two similar Kondo effects acting
      on different sites, each characterized by the energy scale that we call $ T_k $
      . The strong fixed
       point of this problem ($J^*,\ J_z^*\rightarrow +\infty$)
      is stable and corresponds to the well-known Fermi-liquid behaviour: the channel a interacts
      with the half $s_1$ spin of $S_1$ and the channel b interacts with the half
    $s_3$  spin of $S_2$. It remains on each site a $ s=\frac{1}{2} $ "not screened" local moment: $ s_2\in S_1=1,\ s_4\in S_2=1$. It is remarkable
    to notice that these residual  moments
       are totally decoupled from the conduction band and from $s_1$ and $s_3$ respectively.
     Consequently, the conduction electrons are submitted
      to a phase shift $\delta=\frac{\pi}{2}$ induced by the infinite local Kondo coupling. In fact, we
      have solved the case K=0, at a particular solvable limit, where the Kondo
      coupling is not infinite and we could expect that the half-spins $ s_2$ and $s_4$ are not exactly 
      totally decoupled from the conduction band; anyway, the physics is not changed.
      
      \vskip 1cm
      The mean-field treatment appears quite efficient to treat the Kondo problem
      without any interaction K=0 between the two concerned S=1 Kondo impurities and we will
      discuss in the following the non zero K cases.
     
      \subsection{The ferromagnetic coupling (K<0)}
     
      Now we look briefly at a ferromagnetic coupling $K\vec{S_1}\vec{S_2}$, with $ K<0 $. As shown before, an
      RKKY interaction is not expected in this case and K concerns (simply) a direct
      exchange between the two S=1 spins, according to the discussion after eq. (\ref{www}). We just develop
      qualitative arguments
      concerning the phase shift $\delta$ of the conduction electrons induced by the local Kondo effect.
      Indeed, if we consider that the system starts, from K=0, with an infinite
      Kondo coupling ($\delta=\frac{\pi}{2}$) and finally goes towards $K\rightarrow -\infty$ 
      with the same stable Kondo situation ($\delta=\frac{\pi}{2}$), we do not expect 
      (in the area $ K<0 $) any particular critical point where the phase shift of the conduction
      electrons would not be defined. We even expect 
       a Kondo effect of magnitude of $T_k$ for all K<0.
      
      In fact, our mean-field treatment is not well appropriate for the direct
      ferromagnetic K interaction, but our preceding qualitative arguments are sufficient to conclude
      that there is no critical point for K<0, as in the two $s=\frac{1}{2}$ impurity
      case.

       \subsection{The antiferromagnetic coupling (K>0).} 
    
      The most interesting case corresponds to an antiferromagnetic coupling (K>0), because
      in this case it is important to study the absence or existence of a phase
      transition, even a critical point as a function of K, by analogy with the
      two $s=\frac{1}{2}$ impurity case where a sharp phase transition occurs
      for K of order $2T_k$.
      
      However, we will use the mean-field approximation as in the previous K=0
      case and we treat the case of moderate K values, where we can apply only
      a small perturbation from the K=0 results; finally, it is sufficient
      since if a critical point exists, it is certainly not so far from the particular point K=0. 
      
      Thus, we keep
      here $\langle d^{\dag}_1 d^{\dag}_2\rangle=\langle d^{\dag}_3 d^{\dag}_4\rangle=1$
      as previously shown and we consider all the other averages of two operators as small
      quantities.
      
      The mean-field approach gives, therefore:
     
     \begin{equation}
     \label{poi}
      H_{i,\bot}=\frac{K}{2}\{(d^{\dag}_1+d^{\dag}_2)(d_3+d_4)+4(d^{\dag}_1d^{\dag}_4+d^{\dag}_3d^{\dag}_2)+hc\}
      \end{equation}
      \begin{equation}
      \label{pol}
      H_{i,\|}=\frac{K_z}{2}\{\langle d^{\dag}_1 d^{\dag}_4\rangle(d^{\dag}_1 d^{\dag}_4+d^{\dag}_3 d^{\dag}_2)- \sum_{\alpha= {1,2},\ \beta= {3,4}}\langle d^{\dag}_{\alpha}d_{\beta}\rangle d^{\dag}_{\alpha}d_{\beta}+hc\} 
      \end{equation}
      
       We can notice that it is consistent with the mean-field equations
      to consider $\langle d^{\dag}_\alpha d_{\beta}\rangle_{ (\alpha=1,2,\ \beta=3,4)}$ and 
      $\langle d^{\dag}_\alpha d^{\dag}_{\beta}\rangle_{ (\alpha=1,2,\ \beta=3,4)}$ as real. Using the Eqs. [(\ref{poi}),(\ref{pol})], we 
      deduce also that, due to the nonzero K coupling, the antiferromagnetic correlations favorize
     both the pairing mechanism (particle1, particle4) or (particle2, particle3)
      and the binding mechanism of a particle $\alpha (\alpha=1,\ 2)$
      with a hole $\beta (\beta=3,\ 4)$; hence, in contrast to the case  of two 
      $s=\frac{1}{2}$ Kondo impurities\cite{varma}, there is no competition anymore between these 
       two kinds of processes but either a good coexistence; we can add
      that it could be an important argument for the non existence of a phase transition in the area $K\sim T_k$.
     
      Then, we assume that $H_+$ can be treated as a perturbation of $H_{a}$, by
      considering that
       the mean-field symmetries of equations (\ref{kiki}) are preserved. By use of all the previous arguments, the couplings K and $K_z$  renormalize the operators  $\langle d^{\dag}_\alpha d_\beta\rangle_{(\alpha=1,2,\ \beta=3,4)} $ at the same negative value,
       $\langle d^{\dag}_1d^{\dag}_4\rangle =\langle d^{\dag}_3d^{\dag}_2\rangle$
    at a positive constant value, maintain $\langle d^{\dag}_1d^{\dag}_3\rangle$ and $\langle d^{\dag}_2d^{\dag}_4\rangle$
     at zero and make $ H_{+} $ relevant. ${\cal H}$ becomes finally equal to:
      \begin{eqnarray}
      \label{yy}
      {\cal H}&=&{\cal H}_{K=0}+J_1\Psi(d^{\dag}_3+d^{\dag}_4-d_1-d_2)+J_2\Psi(d^{\dag}_2-d^{\dag}_1+d_3-d_4)\\ \nonumber
              &+& \{h_1[(d^{\dag}_1+d^{\dag}_2)(d_3+d_4)+h.c.]+h_2[(d^{\dag}_1 d^{\dag}_4+d^{\dag}_3 d^{\dag}_2)+h.c.]\} 
       \end{eqnarray}
      with the following self-consistent equations:
      \begin{eqnarray}
      \label{ty}
       J_1 &=& 2J\langle d^{\dag}_1d_3\rangle,\qquad J_2 =J\langle d^{\dag}_1d^{\dag}_4\rangle\\ \nonumber
      h_2 &=& -2K+\frac{K_z}{2}\langle d^{\dag}_1d^{\dag}_4\rangle 
      ,\qquad h_1=\frac{K}{2}-\frac{K_z}{2}\langle d^{\dag}_1 d_3\rangle 
      \end{eqnarray}
      
      For an antiferromagnetic coupling, i.e. for positive K and $K_z$ values, one
      can easily show that $J_2$ and $h_1$ are positive, while $J_1$ and $h_2$ are negative; we
      have also $h_2\sim -4h_1$ for small values of $\langle d^{\dag}_1d^{\dag}_4\rangle$
      and $\langle d^{\dag}_1 d_3\rangle$ corresponding to very small values of K. In principle, we
       would have to solve within the mean-field approximation
      the system of self-consistent equations based on the Hamiltonian (\ref{yy})
       and the relations  (\ref{ty}). However, the system is quite tricky to solve
       and we can have already a good insight of the physics in looking simply
       at the solutions obtained within the subspace of operators d and D (or e and
       E) introduced for K=0.
       \\
       Then, we use the equations (\ref{fg1}),(\ref{fg2}),(\ref{fg3}) and(\ref{fg4})
       in order to transform the total Hamiltonian, which becomes:
      \begin{equation}
      \label{trd}
      {\cal H}=H_o^*+H_{res}+H_{coupling}
      \end{equation}
      with
       \begin{equation}
       \label{trs}
       H_{coupling}=(h_1+h_2)(E^{\dag}D+D^{\dag}E)+(h_1-h_2)(e^{\dag}d+d^{\dag}e)+h_1(E^{\dag}d+e^{\dag}D+hc) 
       \end{equation}
       \begin{equation}
       \label{trss}
      H_{res}=J^*\Psi(d^{\dag}-e)+J^*_o(D^+-E)\Psi^{\dag}
      \end{equation}
      and
      \begin{equation}
      \label{trsss}
      J^*_o=J_2-J_1 
     \end{equation}
    
     We have not considered in the equation (\ref{trss}) the small contribution
     $(J_1+J_2)\Psi^{\dag}d^{\dag}$, because the d operator is strongly resonant 
     with the large $J^*$ coupling and $(J_1+J_2)$ is very small; so,
     this very small coupling in  $(J_1+J_2)$ is negligible
     with respect to the very large one in $J^*$ and does not change the physics
     of the problem .
     
     In spite of the peculiar mean-field treatment, the solutions given by the
     above equations yield a good insight on the physics of the two S=1 impurity
     case. $J^*_o$, which is irrelevant for K=0, becomes really relevant for an antiferromagnetic
     coupling. 
     
     The crucial point concerns here the non existence of a 
   critical point or any kind of phase transition as a function of the K parameter, since
 there is no Green function divergent when $\omega\rightarrow 0$ for the considered set of parameters.
 In fact, we have obtained for the two main Green functions (the others vary as
 $K^2$):
 \begin{equation}
 \label{ccc}
 G_{dd^{\dag}}(\omega)=\frac{-i(\omega\pm\Gamma_k)\{{\omega}^2\pm\Gamma_k\omega
 +h_1^2+{(h_1-h_2)}^2\}}{{({\omega}^2\pm\Gamma_k\omega+h_1^2+{(h_1-h_2)}^2)}^2-4h_1^2(h_1-h_2)^2}
 \end{equation}
 and
 \begin{equation}
 \label{ccd}
 G_{DD^{\dag}}(\omega)=\frac{-i(\omega\pm\Gamma_o^*)\{{\omega}^2\pm\Gamma_o^*\omega
 +h_1^2+{(h_1+h_2)}^2\}}{{({\omega}^2\pm\Gamma_o^*\omega+h_1^2+{(h_1+h_2)}^2)}^2-4h_1^2(h_1+h_2)^2}
 \end{equation}
where $\omega=(2n+1)\pi/\beta$ is a fermionic Matsubara frequency $\Gamma_k=\rho J^{*2}$ and $\Gamma_o=\rho J_o^{*2}<<\Gamma_k$
are respectively the widths of the d and D (respectively the e and E) impurity levels, $\Gamma_o^*=\Gamma_o+iE_D$
and $\Gamma_k$ is of order $T_k$. In Eqs. (\ref{ccc}) and (\ref{ccd}), the upper and
lower sign corresponds respectively to the case of positive and negative $\omega$. Indeed, 
the different Green functions do not develop a pole at $\omega=0$, whatever the values
of $\Gamma_k$ and $\Gamma_o$ are and we do not expect any critical point at low temperatures. Especially, we expect that the
staggered susceptibility $\chi_s$ does not diverge\cite{varma}.

Simply, when K and $K_z$ are small, we can neglect the terms in $h_1^2$ and 
$h_2^2$ and therefore, write the two (main) Green functions:
\begin{eqnarray}
G_{dd^{\dag}}(\omega)&=&\frac{1}{i\omega\pm i\Gamma_k}=G_{ee^{\dag}}\\ \nonumber
G_{DD^{\dag}}(\omega)&=&\frac{1}{i\omega\pm i\Gamma_o^*}=G_{EE^{\dag}}
\end{eqnarray}
The influence of K appears mainly through the magnitude $\Gamma_o$. So, when K
is small, it appears two cohabiting species of quasiparticles: heavy quasiparticles
with an electronic specific heat constant $C/T=\Gamma_k^{-1}(K)=\chi$ and quasi-free
electrons which lead to the main RKKY interaction between the non-screened half-spins, namely
$s_2$ ans $s_4$ (already introduced for K=0) and generated by a small ferromagnetic Kondo coupling due to the
Pauli principle. In fact, other marginal RKKY interactions also exist, which
 couple all the half-spins and which, especially guarantee exact physics in the strong K-coupling limit; however, they
  are not really relevant, for small values of K and can be forgotten.
 
Indeed, all these conclusions
are done, at a particular solvable point and we can not be exactly sure that 
they remain true for any value of J; nevertheless, we think that these results are 
physically correct and then, the fixed point of the coupling J has to 
decrease with K. Precisely, the dominant RKKY interaction (between $s_2$ ans $s_4$)
tends to suppress
the critical point, obtained with the two s=1/2 Kondo impurity-model and yields
both a Kondo effect and magnetism, for small positive K values.

    \vskip 1cm
    \section{Conclusion}
    \vskip 1cm
    In this paper, we have presented an explicit study of the problem of two S=1 magnetic
    impurities interacting with a conduction band and coupled via an interimpurity  coupling $K\vec{S_1}.\vec{S_2}$. There 
    is no quantum critical
    point, even no phase transition in the phase diagram and this last point is very important because it
    shows a behaviour completely different from that of the regular two screened $s=\frac{1}{2}$ impurity Kondo
    model. In fact, a {\bf smooth} cross-over separates a ``one-underscreened-Kondo-impurity'' like  phase from an antiferromagnetic and non-Kondo phase. In
    particular, it leads that $\delta=\arctan(J^*)$ and $\frac{\partial}{\partial K}\langle\vec{S_1}\vec{S_2}\rangle$
    vary continuously with K, for all the real values of this parameter.  
    
    We have obtained, for a positive and small K value, an asymmetric situation with
    a strong Kondo effect for the spins $s_1$ and $s_3$, a weak Kondo effect for the spins
    $s_2$ and $s_4$ and finally a RKKY interaction between $s_2$ and $s_4$. This can lead
    to a coexistence between a Kondo effect leading to strong spin fluctuations on
    one side and an indirect RKKY interaction. Finally, with only two spins, a true 
    magnetic order and a really broken SU(2) spin symmetry could not occur but
    it is encouraging to yield, even in this particular case, a coexistence between a heavy-fermion character 
    and (special) magnetism.
    
    Thus, the
    case of a moderate and antiferromagnetic K coupling can account for the physics
    of Uranium compounds, such as $UPt_3$, where both a heavy-fermion behaviour
    and some kind of long-range magnetic order exist at low temperatures. In
    any underscreened  Kondo lattice model, the presence of magnetism is expected
    but much remains to be understood concerning the magnetic length of the
    intersite antiferromagnetic fluctuations or more generally concerning
    the tiny magnetic moment which characterizes the magnetic character
    of $UPt_3$, as already noticed by Coleman et al. \cite{col}. Finally, a more complete
     explanation of the properties of compounds such as $UPt_3$, based on a
    non-Abelian treatment of the underscreened Kondo lattice, is presently studied\cite{kar}.
     
     \acknowledgments

    One of us (K.L.H.) would like to thank Thierry Giamarchi for interesting discussions and remarks.

      \end{document}